\documentclass[a4paper]{PoS}

\title{Classical novae vs cataclysmic variables}

\ShortTitle{CNe vs CVs}

\author{\speaker{C.Tappert}\\
 Instituto de F\'isica y Astronom\'ia, Universidad de Valpara\'iso, 
 Valpara\'iso, Chile\\
 E-mail: \email{claus.tappert@uv.cl}}

\author{I. Fuentes-Morales\\
 Instituto de F\'isica y Astronom\'ia, Universidad de Valpara\'iso, 
 Valpara\'iso, Chile\\
 E-mail: \email{irma.fuentes@uv.cl}}

\author{E. Puebla\\
 Instituto de F\'isica, Pontificia Universidad Cat\'olica de Valpara\'iso, 
 Valpara\'iso, Chile\\
 E-mail: \email{caro\_lita21@hotmail.com}}

\author{A. Ederoclite\\
 Centro de Estudios de F\'isica del Cosmos de Arag\'on, Teruel, Spain\\
 E-mail: \email{aederocl@cefca.es}}

\author{L. Schmidtobreick\\
 European Southern Observatory, Santiago, Chile\\
 E-mail: \email{lschmidt@eso.org}}

\author{N. Vogt\\
 Instituto de F\'isica y Astronom\'ia, Universidad de Valpara\'iso, 
 Valpara\'iso, Chile\\
 E-mail: \email{nikolaus.vogt@uv.cl}}

\abstract{We present a preliminary comparison of the post-nova population with
that of general cataclysmic variables (CVs). We particularly focus on the 
mass-transfer rate and its potential relation to the nova eruption. We find 
that the known post-nova sample exclusively consists of high mass-transfer CVs,
but that this is more likely to be due to the shorter recurrent time for those
systems, rather than the mass-transfer rate being affected by the eruption.
Nevertheless, we find evidence for such an effect for specific post-novae,
and that it is potentially related to the binary separation and to presence
or absence of an accretion disc.}

\FullConference{The Golden Age of Cataclysmic Variables and Related Objects - III, Golden2015 \\
		7-12 September 2015\\
		Palermo, Italy}

\begin{document}

\section{Introduction}

A statistical comparison of CVs that have undergone a nova eruption (hereafter
called `post-novae') and those that are not known to have done that (hereafter
referred to as `non-novae') should indicate the physical parameters that are 
important for the nova eruption as well as the effect of the nova eruption on 
the parameters. For this, both samples need to be of a size that allows for a 
statistical significant comparison. However, for most questions, the sample of 
the post-novae does not fulfil this requirement. On the one hand this is due
to the inherent small number of novae. These are comparatively rare events,
and the number of registered nova eruptions that occurred before 1986 amounts 
to little more than 220 (compared to the over 3000 known CVs, e.g.
\cite{downesetal05-1}). On the other hand, the usable sample is even much
smaller, because more than half of those reported novae still lack the
identification of the post-nova system. Over the past few years, our group has
conducted a survey of previously unidentified post-novae in order to
recover and classify them \cite{tappertetal12-1,tappertetal14-1,%
tappertetal15-2,ederoclite,fuentesmorales}. In spite of these efforts, an
analysis of the sample is still plagued by small-number statistics.
Nevertheless, while being aware of this difficulty, in the present,
preliminary, study we will investigate the properties of the available
sample of post-novae to search for a possible connection to CV parameters.

We here concentrate on the possibly most important parameter for the nova
eruption, the mass-transfer rate $\dot{M}$. It is immediately clear that
a high mass-transfer rate will result in a shorter nova eruption recurrence
time, as less time is needed to pile up the critical mass on the white
dwarf. Additionally, it has been shown that for lower $\dot{M}$
a larger amount of mass needs to be accumulated to trigger the nova
eruption, thus augmenting this effect 
\cite{townsley+bildsten04-1,townsley+bildsten05-1}. The influence of $\dot{M}$
on the nova eruption thus appears to be obvious, but in which way and amount
does the latter affect the former? In the hibernation model, the 
eruption-heated white dwarf irradiates the donor star and causes a phase of 
enhanced $\dot{M}$ \cite{sharaetal86-1,prialnik+shara86-1}. As the
white dwarf cools down, $\dot{M}$ decreases, potentially including a prolonged
state of a detached system. This model describes a cyclic behaviour from the
nova eruption over a high $\dot{M}$ CV 
(with $\dot{M} \sim 10^{-9}~\mathrm{M}_\odot~\mathrm{yr}^{-1}$; 
e.g.~\cite{townsley+gaensicke09-1}) to a low $\dot{M}$ CV 
($\dot{M} \sim 10^{-11}-10^{-10}~\mathrm{M}_\odot~\mathrm{yr}^{-1}$) and back
(see also \cite{vogt82-2}), with the necessary ingredient that \emph{all}
post-novae have high $\dot{M}$. The opposite possibility is that the novae
originate from the individual pools of high or low $\dot{M}$ CVs, i.e.~either
pre- and post-nova are high $\dot{M}$ CVs or they are low  $\dot{M}$ CVs. In
other words, the nova eruption would have negligible influence on the
post-nova's $\dot{M}$. Also for this model one would expect the large majority
of post-novae to have high $\dot{M}$, because of the much shorter
eruption recurrence time, but there should exist a considerable number of
low $\dot{M}$ post-novae, since low $\dot{M}$ CVs represent between 70 and
99 per cent of the CV population \cite{stehleetal97-1,kniggeetal11-1}. Finally,
there is the intermediate case, that $\dot{M}$ is affected by the nova eruption
only for a certain subsample of CVs, e.g.~distinguished by containing a 
low-mass white dwarf or having a short orbital period 
\cite{pattersonetal13-2,nelemansetal15-1,schreiberetal16-1}.

\section{Mass-transfer rates in post-novae}

The quantitative determination of $\dot{M}$ is difficult, depending on
accretion disc models and distance determinations. In the following we list
qualitative $\dot{M}$ indicators and the potential pitfalls associated with
them.

\subsection{Spectral indicators}

High $\dot{M}$ discs are optically thick, resulting in weak emission lines
superposed on a blue continuum. For a steady-state disc, the spectral energy
distribution (SED) $F(\lambda)$ for wavelength $\lambda$ follows an exponential 
law as $F(\lambda) \propto \lambda^{-\alpha}$ with $\alpha = 7/3$ 
\cite{lyndenbell69-3}. For lower $\dot{M}$, the slope decreases, and once
the stellar components begin to significantly contribute to the SED in the 
visual range, it no longer can be described by the form above. Most novae are 
located in the Galactic disc \cite{ederocliteetal11-1}, and thus the 
determination of the intrinsic SED is hampered by interstellar reddening. 
While some attempts have been undertaken to correct for this effect using 
large-scale reddening maps (e.g.~\cite{schlafly+finkbeiner11-1} for
\cite{tappertetal12-1,tappertetal14-1,tappertetal15-2}), this really needs
data with high spectral resolution in order to measure interstellar absorption
lines. 

The strength of the emission lines from the disc is inversely related to
the disc brightness and thus to $\dot{M}$ \cite{patterson84-1}. However,
in CVs the lines frequently include contribution from additional emission 
sources, e.g.~from the bright spot. In post-novae, specifically, irradiation
of the donor star by the hot white dwarf can result in additional line
emission (e.g., \cite{tappertetal13-1}). Furthermore, emission from the
nova shell will contribute to the optical spectrum for a considerable amount
of time after the nova eruption. While we restrict our study to post-novae
that are `older' than 30 years in order to minimize this effect, there are
some systems that show significant contribution from the shell much longer
than that \cite{fuentesmorales}. 

The problems with using the emission line strengths as an indicator for 
$\dot{M}$ are best illustrated using the work from Zhao \& McClintock
on the eclipsing old nova OY Ara (N Ara 1910) \cite{zhao+mcclintock97-1}. 
This object is one of the post-novae with the strongest Balmer emission lines,
the equivalent width of H$\beta$ amounting to $W_\mathrm{H\beta} \sim 30$ 
{\AA}. Using Patterson's relation \cite{patterson84-1}, its disc brightness 
results to $\sim$8.5 mag, placing it among the high $\dot{M}$ dwarf novae like 
AH Her and Z Cam. However, the time-resolved spectra show a strong dependency 
of the emission line strengths on the orbital phase, with maximum strength
corresponding to orbital phase $\sim$0.8. It can thus be assumed that emission
from the bright spot contributes significantly to the emission line. In fact,
the eruption parameters lead Zhao \& McClintock to estimate the disc 
brightness to 3.6 mag, a value that should correspond to 
$W_\mathrm{H\beta} \sim 5$ {\AA}.

\subsection{Eruption amplitude}

\begin{figure}
\includegraphics[width=.49\textwidth]{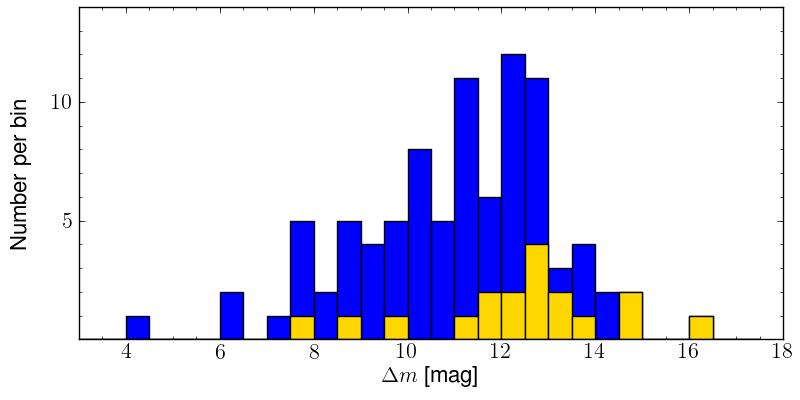}
\includegraphics[width=.49\textwidth]{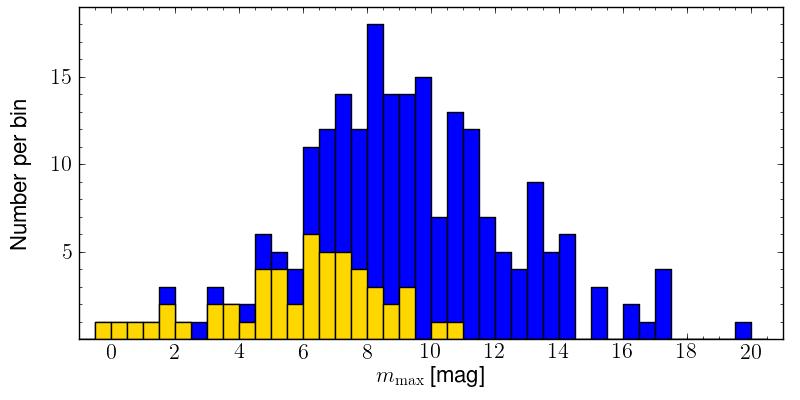}
\caption{Left: Histogram of the eruption amplitudes. The yellow (bright)
histogram corresponds to likely magnetic post-novae. Right: Distribution of
the reported maximum eruption magnitudes. Here, the blue (dark) histogram
indicates the yet unidentified post-novae, while the yellow (bright) one
refers to the known systems. In both histograms the samples
consists of pre-1986 novae.}
\label{ct_amphis}
\end{figure}

The brightness of accretion discs in CVs spreads over $\sim$9 mag 
\cite{patterson84-1}, while the brightness at maximum for novae ranges from
$\sim$ $-6$ to $-10$ mag \cite{dellavalle+livio95-1,kasliwaletal11-2,%
hachisu+kato15-3}. Thus, the spread in an eruption amplitude 
$\Delta m = m_\mathrm{pn}-m_\mathrm{max}$, i.e.~defined as the difference 
between the brightness at maximum $m_\mathrm{max}$ and the brightness of the 
post-nova $m_\mathrm{pn}$, will mainly reflect the spread in the quiescence
magnitude of the post-novae. Essentially, low $\dot{M}$ post-novae will
have larger $\Delta m$ than high $\dot{M}$ systems. Potential pitfalls are here
that the post-nova brightness depends on the system inclination (edge-on 
systems are fainter), and that the nova had not been observed at its actual
maximum brightness, resulting in a fainter recorded $m_\mathrm{max}$.

The faint end of the absolute brightness of novae at maximum amounts to
roughly $-$6.5 mag in the visual range \cite{yaronetal05-1}, while
low $\dot{M}$ CVs (dwarf novae) typically have absolute quiescence brightness 
$\ge$8.0 mag. This yields a lower limit of $\Delta m \sim$14.5 mag for the 
amplitude of a post-nova to reach even only the high $\dot{M}$ end of the 
dwarf novae. The distribution of the eruption amplitudes of the pre-1986 novae 
(left plot in Fig.~\ref{ct_amphis}) shows that there is no currently
known non-magnetic system that passes this limit. However, the shape of
the distribution, in particular the abrupt drop in numbers at $\Delta m = 13$
mag, indicates the presence of an observational bias. It appears reasonable
to define a post-nova with a quiescence brightness of about 22 mag as `hard to 
detect'. Thus, all novae with apparent maximum brightness 
$m_\mathrm{max} > 7.5$ mag and $\Delta m \sim$14.5 mag fall into this category.
From the right plot in Fig.~\ref{ct_amphis} we see that this concerns only
37 per cent of the known post-novae, but about 87 per cent of the yet
unidentified systems. It is thus possible that the latter sample contains a 
significant number of large amplitude novae.

\subsection{The orbital period distribution}

\begin{figure}
\includegraphics[width=0.6\textwidth]{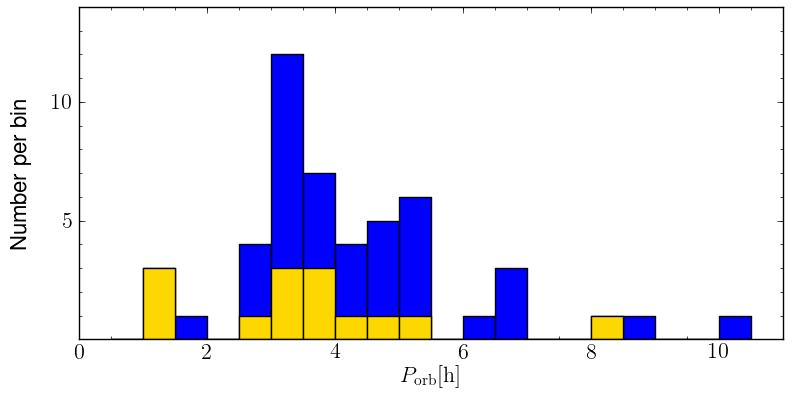}
\caption{The distribution of the orbital periods in the pre-1986 post-novae
(from \cite{ritter+kolb03-1} and \cite{tappertetal13-2}).
The yellow (bright) histogram indicates likely magnetic systems.}
\label{ct_phis}
\end{figure}

Although there exists a significant spread in mass-transfer rates especially
for CVs with orbital periods $P_\mathrm{orb} > 3$ h, the distribution of the
periods can be used as a rough indicator for $\dot{M}$. In particular,
systems below the period gap have $\dot{M}$ typically 1--2 orders
of magnitude lower than above the gap, and CVs with the highest
$\dot{M}$ lie in the 3--4 h range \cite{townsley+gaensicke09-1}. The period
distribution of post-novae is significantly different from that of
non-novae (Fig.~\ref{ct_phis}; see also \cite{ritter+kolb03-1,%
townsley+bildsten05-1}). First, while the vast majority of non-novae lie
below the period gap, less than 10 per cent of post-novae have 
$P_\mathrm{orb} < 2$ h. Secondly, the distribution of the post-novae
shows a distinctive peak in the 3--4 h range. These two features indicate
that the post-nova population is dominated by intrinsically high $\dot{M}$ 
systems rather than high $\dot{M}$ being a consequence of the nova eruption.

\subsection{Long-term light curves}

\begin{figure}
\includegraphics[width=.49\textwidth]{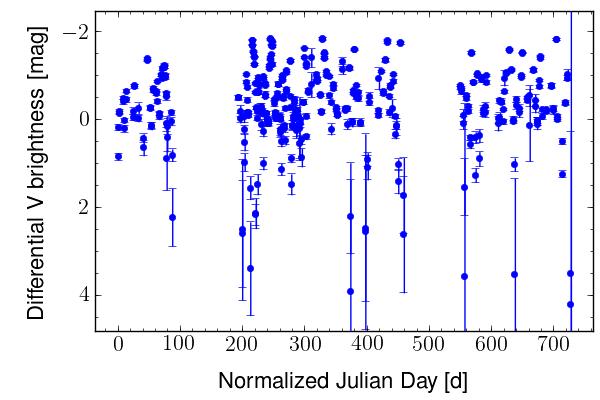}
\includegraphics[width=.49\textwidth]{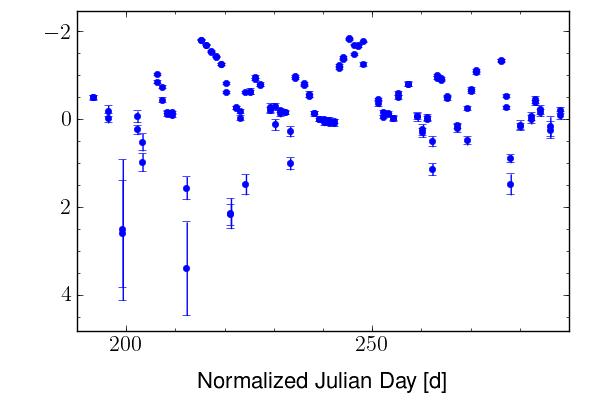}
\caption{Long-term light curve of the eclipsing post-nova V728 Sco. The x-axis 
indicates the time in days after the first observation, the y-axis represents
the brightness in $V$ normalized to the quiescence out-of-eclipse value.
The right plot presents a close-up of the time range just after
the first observational gap.}
\label{ct_v728}
\end{figure}

The ultimate test for the mass-transfer rate is the long-term behaviour. Low  
$\dot{M}$ discs are not in a steady state, but subject to outburst behaviour,
with frequency, amplitude and duration being further indicative of $\dot{M}$
\cite{osaki05-1,ichikawa+osaki94-1}. Very few post-novae are known to
show disc outbursts, the primary example being V446 Her (Nova Her 1960)
\cite{honeycuttetal95-1,honeycuttetal98-2,honeycuttetal11-1}. In 
Fig.~\ref{ct_v728} we present preliminary long-term data of the eclipsing 
post-nova V728 Sco (Nova Sco 1862) taken on the 1.3 m SMARTS telescope at 
Cerro Tololo Inter-American Observatory (Chile) between 2013 August 1st and 
2015 July 29th. While the data still have to be corrected for orbital
variations, they clearly show variations that strongly resemble disc outbursts.
Like in V446 Her, the outbursts are more frequent and have a lower amplitude
than those often observed in dwarf novae. Schreiber et 
al.~\cite{schreiberetal00-2} 
explained these differences with the presence of a hot ionized inner disc
that is the result of irradiation by the eruption-heated white dwarf. 
Analysis of the eclipse profile in V728 Sco indeed provides evidence for
the presence of such a disc \cite{tappertetal13-1}.

\subsection{Brightness before and after the eruption}

\begin{figure}
\includegraphics[width=.49\textwidth]{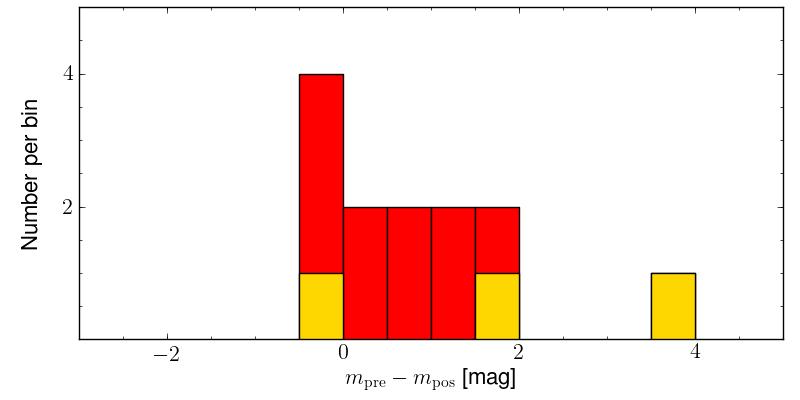}
\includegraphics[width=.49\textwidth]{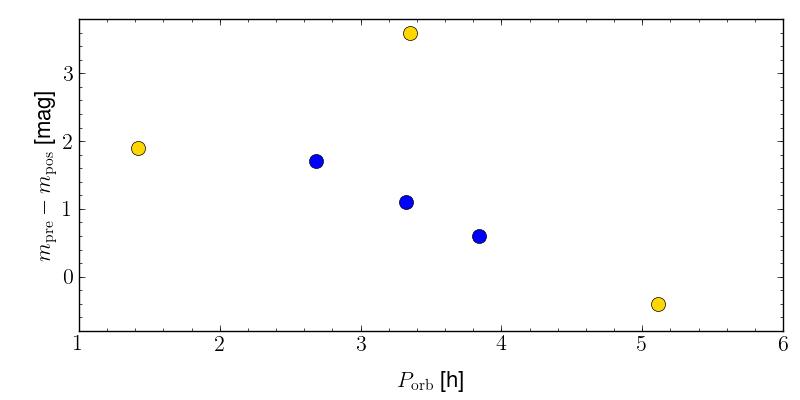}
\caption{Left: Histogram of the difference between the pre- and post-nova
brightness in pre-1986 novae. Right: The difference as a function of the 
orbital period. In both plots, the yellow (bright) data mark likely magnetic
systems. The brightness data were taken from \cite{collazzietal09-1}.}
\label{ct_mpre}
\end{figure}

A comparison of the brightness of the pre-nova $m_\mathrm{pre}$ with that of 
the post-nova $m_\mathrm{pos}$ will yield the most immediate indication
of the influence of the nova eruption on the CV. However, data on pre-novae
are still scarce, with the only comprehensive study being that of Collazzi
et al.~\cite{collazzietal09-1}. The right plot in Fig.~\ref{ct_mpre} shows
a histogram of these data for our sample of pre-1986 novae, amounting to just
13 systems. The result is not unambiguous. For about half of the systems the 
difference is smaller than 0.5 mag, i.e.~there appears to be no significant 
effect of the eruption. Five objects have differences $\ge$1 mag,
among them the discless post-nova V1500 Cyg with the largest value of
almost 4 mag. The left plot of Fig.~\ref{ct_mpre} indicates a possible
inverse relation of the brightness difference with the orbital period. The fact
that V1500 Cyg as the only discless system represents the only strong
exception from this apparent relation could hint that there are two important
parameters that soften the impact of the nova eruption on the CV: the binary
separation and the presence of an accretion disc. We emphasize that this is
based on very few data points and that a more comprehensive study is needed
that also takes into account further parameters, e.g.~the time that has passed
since the nova eruption.

\section{Summary}

We have conducted a preliminary analysis of the available post-nova sample
for novae that erupted before 1986. We have especially explored the
potential effect of the nova eruption on the mass-transfer rate $\dot{M}$
of the post-nova. We find that, although the post-nova sample appears to
exclusively consist of high $\dot{M}$ systems, that this is potentially
due to an observational bias towards low eruption amplitude novae. Furthermore,
the orbital period distribution indicates that the post-nova population is
dominated by intrinsically high $\dot{M}$ CVs with short eruption recurrence
times. In other words, it is the high $\dot{M}$ that causes the nova, rather
than the other way round. Still, the comparison of the pre- and post-nova
brightnesses shows that for a significant fraction of CVs brightness and 
thus likely $\dot{M}$ are affected by the nova eruption, most probably due to
irradiation by the heated white dwarf, as predicted and required by the
hibernation scenario. It appears that the condition of
such irradiation being effective is a combination of a small binary separation
and the absence of an accretion disc. Thus, short-period magnetic systems
should be the most affected by the nova eruption, while the disc appears
to block a significant amount of irradiation, perhaps instead resulting
in the formation of an ionized inner disc like suspected in V728 Sco. 

We point out that the present small size of the post-nova sample potentially
affects the result of our analysis, and emphasize the need for more data.
The recent results on novae from the Optical Gravitational Lensing Experiment 
(OGLE, \cite{mrozetal15-3}) show that present and future long-term monitoring 
of the sky will provide a valuable advance especially with respect to the 
characterization of the pre-nova state.

\acknowledgments
This research was supported by FONDECYT Regular grant 1120338 and the
Centro de Astrof\'isica de Valpara\'iso (CT, IFM and NV). AE acknowledges 
support by the Spanish Plan Nacional de Astrononom\'ia y Astrof\'isica under 
grant AYA2012-30789.

\bigskip
\bigskip
\noindent {\bf DISCUSSION}

\bigskip
\noindent {\bf DEANNE COPPEJANS:} 
In the Schreiber et al.~(2000) model, the stunted outburst are attributed to
an irradiated inner disc. Is the idea that the irradiated inner disc is in
a high-state constantly, so that quiescence is brighter and the outburst
amplitude is smaller, or is it that the inner disc is dissipated and so 
can't take part in the outburst?

\bigskip
\noindent {\bf CLAUS TAPPERT:} The idea is that the inner disc is in an
ionized configuration due to irradiation by the hot white dwarf. As such, it
is stable and does not participate in the disc outburst. So, the part of
the disc being affected by the outburst is smaller, thus accounting for the
smaller outburst amplitude. Additionally, the cooling wave cannot traverse
through the whole disc, resulting in a higher outburst frequency.
The model predicts that, as the white dwarf cools down, the amplitude should
increase, and the frequency should decrease with time.
\end{document}